\documentstyle[preprint,prl,aps,epsf]{revtex}

\begin{document}
\draft
\title{Collective excitations at the boundary of a 4D quantum Hall droplet}
\author{Jiangping Hu and Shou-Cheng Zhang}
\address{
Department of Physics, Stanford University, Stanford, CA 94305
}


\maketitle
\begin{abstract}
In this work we investigate collective excitations at the boundary
of a recently constructed 4D quantum Hall state. Local bosonic
operators for creating these collective excitations can be
constructed explicitly. Massless relativistic wave equations with
helicity $S$ can be derived exactly for these operators from their
Heisenberg equation of motion. For the $S=1$ and $S=2$ cases these
equations reduce to the free Maxwell and linearized Einstein
equation respectively. These collective excitations can be
interpreted as hydrodynamical modes at the boundary of the 4D QHE
droplet. Outstanding issues are critically discussed.

\end{abstract}
\newpage

\narrowtext
The two dimensional quantum Hall liquid state\cite{laughlin} provided us with
much insight into the novel and surprising organization principles of matter.
Recently a four dimensional quantum Hall liquid state
has been constructed\cite{4dqhe}.
The system consists of non-relativistic fermions moving on
a four dimensional sphere $S^4$, interacting with a background $SU(2)$
gauge field. The gauge field is created by the $SU(2)$
monopole\cite{yang1,yang2}, which can be obtained by a conformal
transformation\cite{jackiw,polyakov} of the Belavin-Polyakov-Schwartz-Tyupkin
instanton\cite{instanton} in
the four dimensional Euclidean space. The fermions are in the
$I$th representation of the $SU(2)$ gauge field, and all
eigenstates form irreducible representations of the $SO(5)$ group,
which is the isometry group of the four sphere. In the lowest
$SO(5)$, or generalized Landau level, the ground state degeneracy
is $D(p)=\frac{1}{6}(p+1)(p+2)(p+3)$, with $p=2I$. The simplest
many-body system to consider is when the filling factor
$\nu=N/D(p)=1$.

\begin{figure*}[h]
\centerline{\epsfysize=6.0cm \epsfbox{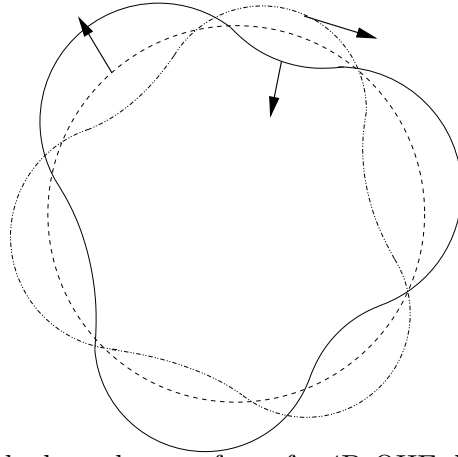} }
 \caption{An illustration of the boundary surface of a 4D QHE
 droplet. There is one surface for every isospin direction,
 indicated by an arrow.
. }
 \label{surface_fig}
\end{figure*}

A surface boundary can be introduced in a fashion similar to the edge states of
the 2D QHE\cite{halperin,wen,stone,fisher},
by applying a confining potential $V(x_5)$, which confines the
fermions in a region close to the north pole. The resulting 3D surface of the
4D QHE droplet can be visualized in a way similar to a fermi surface, but
with $2I+1$ distinct copies in real space, one for each isospin directions
of the underlying fermions. (See Fig. \ref{surface_fig} for an illustration).
Elementary excitations of this 4DQHE droplet can be described in different
ways. In principle, the fermion operators offer a full description of the
excitations. On the other hand, we are also interested
in finding a particular class of particle-hole excitations, defined by {\it local}
bosonic particle-hole operators, which describe hydrodynamical distortions of the
droplet surface. Since there are $2I+1$ different copies of the droplet surface,
we expect $2I+1$ different hydrodynamical modes. In the 2D QHE, these two descriptions
are fully equivalent to each other, thanks to the bosonization in 1+1 dimensions.
In our case, because of the higher dimensionality, there are also other,
fermionic excitations besides the hydrodynamical modes. A key finding of our
work is that these collective excitations can be created by {\it local} bosonic
operators which obey massless relativistic wave equations with
helicity $0\leq S \leq I$. In the case of $S=1$ and $S=2$, these relativistic wave equations
are exactly free Maxwell and linearized Einstein equations respectively.

However, it should be warned that these relativistic bosons are non-interacting
at the current level, and their interactions with each other and with the
fermionic part of the 4D QHE droplet need to be carefully studied in
the future. On the other hand, our experience with the zero sound mode in
the ordinary fermi liquid teaches us that the hydrodynamical modes are usually
highly robust against both interactions and reorganization of the fermi
sea. For example, when a superconducting gap opens up in the single particle
fermi spectrum, the collective sound mode is completely unaffected by this
dramatic reorganization. For this reason, we shall concentrate on the collective
modes first, and address the fermionic parts of the excitations at a later
stage. In fact, we believe that there are many possible quantum liquid phases
in this model, and only after a proper reorganization of the
fermi spectrum, a fully relativistic theory can be obtained in the low
energy sector.

Our findings might be important for the idea that massless relativistic
particles can be composite, rather than elementary. If one starts
from a relativistic system, the Weinberg-Witten
theorem\cite{weinberg-witten} states that it is not possible to describe
a higher helicity particle with $S>1$ as a composite object. In
our case, we actually start from non-interacting, non-relativistic
fermions, where this theorem does not apply. However, it is
counter to our intuition that one can form a bound state out of
non-interacting particles. The basic argument runs as follows. Let
us consider a particle-hole operator
\begin{eqnarray}
\rho_{q,k}=c^\dagger_{q/2+k}c_{-q/2+k}=\sum_{x,y}e^{i(q/2+k)x}e^{-i(-q/2+k)y}
c^\dagger_x c_y=\sum_{Z,z}e^{iqZ}e^{ikz}c^\dagger_{Z+z/2} c_{Z-z/2}
\label{rho}
\end{eqnarray}
where $q$ and $k$ are the center-of-mass and relative momenta respectively.
$x$ and $y$ label the coordinates of the particle and the hole, while $Z=(x+y)/2$
and $z=x-y$ label their center-of-mass and relative positions.
This operator is an exact eigen-operator of the non-interacting Hamiltonian,
which satisfies the equation of motion
$\partial_t \rho_{q,k}=(\epsilon_{q/2+k}+\epsilon_{-q/2+k}) \rho_{q,k}$,
where $\epsilon(q)$ is the energy of a plane wave state. The problem is that
this operator is not {\it local}. From (\ref{rho}) one can see that this state
is constructed from a particle-hole pair with all relative positions $z$, each with
equal weight $|e^{ikz}|=1$. Therefore, this operator does not create a
{\it local}, or particle like excitation in the system. One can form a local operator,
$\rho(Z)=e^{iqZ}\sum_{k}\rho_{q,k}$, but the problem is that this operator
does not propagate at a well-defined energy, since different $k$ components carry
different energies. Therefore, if one initially creates a local excitation
at $Z$ by using the $\rho(Z)$ operator, the wave packet will spread over time,
until all energies are dissipated. For this reason it is impossible to
construct any local bosonic operators which obey well defined wave equations.
It is of course possible to construct such local operators with well defined
dispersion in an interacting system, but these operators in general do not have
well defined, non-trivial helicities. Throughout this work, we use the conventional
definition of the helicity as ${\bf S\cdot q/|q|}$, where ${\bf S}$ is the spin
rotation operator and ${\bf q}$ is the linear momentum. A special form of the
spin orbit coupling is required to construct states with well defined
helicities. The special form of the spin orbit coupling is the central
point of this paper.

\begin{figure*}[h]
\centerline{\epsfysize=6.0cm \epsfbox{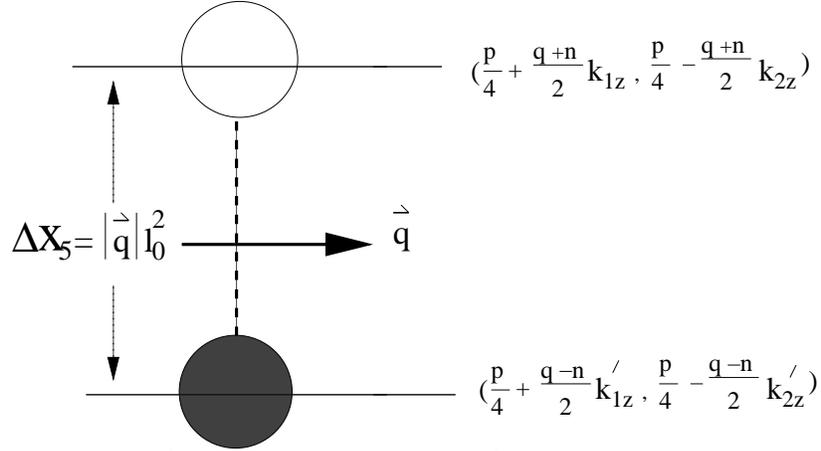} }
 \caption{An illustration of the extremal dipole configuration. For
 a given center of mass momentum $q$, the dipole separation in the
 extra dimension is given by $ql_0^2$. $SO(4)$ quantum numbers of the
 particle and the hole are also indicated. }
 \label{dipole_fig}
\end{figure*}

The above argument breaks down if the fermionic states are
not ordinary plane wave states, but eigenstates of the $SU(2)$
magnetic translation group,
\begin{eqnarray}
[X_\mu,X_\nu]= 4 i l_0^2 \eta_{\mu\nu}^i \frac{I_i}{I}
\label{non-commutation}
\end{eqnarray}
which is the central algebraic structure identified in
\cite{4dqhe}. Because of this structure, one can construct a class
of {\it extremal dipole operators}, which are {\it localized} to
the maximal possible extent in the three dimensional relative
coordinates $X_1,X_2,X_3$ of the particle and the hole, but {\it
stretched} to the maximal possible extent in the relative
coordinate in the fourth dimension $X_4$, see Fig. \ref{dipole_fig} for
an illustration. These dipoles are closely
analogous to the particle-hole dipoles in the 2D
QHE\cite{kallin,stone,haldane,dhlee}. Even though the particles are not
mutually interacting, a bound state can be formed because the force
due to the confining potential is counter balanced by the Lorentz
force when the dipole is propagating. Edge states in the 2D integer
QHE can be understood in terms of this dipole picture\cite{stone}.
In that case, the dipole description is fully equivalent to the
hydrodynamical description\cite{wen}.
In our case, only the extremal dipole states correspond
to collective shape distortion at the boundary, there are other
fermionic excitations are the boundary as well. We shall prove a
mathematically precise result which states that these {\it local}
operators obey exactly the massless relativistic bosonic wave
equations with helicity $S$. For the $S=0,1,2$ cases, these
equations reduce to the Klein-Gordon, the free Maxwell and the
linearized Einstein equations respectively. In this case, the nontrivial
helicities are obtained because of the underlying spin-orbit
coupling imposed by the background monopole field. At this stage, this
result should be considered as a purely mathematical statement,
that these free relativistic wave equations can be derived from a
single non-relativistic Schroedinger equation. On the other hand,
in view of the difficulties with other approaches mentioned above,
we view this as an interesting and remarkable result. It also reveals the deep
algebraic structure encoded in equation (\ref{non-commutation}),
which we are just beginning to understand. The full physical
significance of this kinematic result can only be understood when
interactions are fully considered, and we will comment on
outstanding problems towards the end of this paper. In this paper,
we shall follow the notations and conventions of ref.\cite{4dqhe}.

We parameterize the four sphere $S^4$ by the following coordinate system
\begin{eqnarray}
x_1 &=& sin\theta sin{\frac{\beta}{2}}sin(\alpha-\gamma) \\
x_2 &=& -sin\theta sin{\frac{\beta}{2}}cos(\alpha-\gamma) \\
x_3 &=& -sin\theta cos{\frac{\beta}{2}}sin(\alpha+\gamma) \\
x_4 &=& sin\theta cos{\frac{\beta}{2}}cos(\alpha+\gamma) \\
x_5 &=& cos\theta,
\end{eqnarray}
where $\theta, \beta \in [0,\pi)$ and $\alpha,\gamma \in
[0,2\pi)$. The direction of the isospin is specified by
$\alpha_I,\beta_I$ and $\gamma_I$. In the lowest $SO(5)$ level,
the kinetic energy is constant, and can be taken to zero. The
Hamiltonian is simply given by the confining potential
\begin{eqnarray}
H=cR\sum_{<m>}(m_3+m_4-m_1-m_2)c^+_{<m>}c_{<m>}
\label{H}
\end{eqnarray}
where $<m>=<m_1,m_2,m_3,m_4>,\ \  \sum_{\alpha} m_{\alpha}=p$,
and $c^\dagger_{<m>}$ creates a state in the lowest $SO(5)$
level. The single particle wave function for the $<m>$ state
is given by:
\begin{eqnarray}
\Psi^{k_1,k_2,I'_z}_{k_{1z}k_{2z}}(\theta,\alpha,\beta,\gamma,\alpha_I,\beta_I,\gamma_I)
=f_{k_1,k_2}(\theta)
G^{k_1,k_2,I'_z}_{k_{1z},k_{2z}}(\alpha,\beta,\gamma,\alpha_I,\beta_I,\gamma_I),
\ \ k_1+k_2=I,
 \label{low}
\end{eqnarray}
where
\begin{eqnarray}
f_{k_1,k_2}(\theta)=(-1)^{k_1-k_2}(2I^2+3I+1)\left[
\frac{(I+1)(2I)!}{(2k_1)!(2k_2+1)!} \right ]^{\frac{1}{2}}
\frac{(1+cos\theta)^{2I+\frac{3}{2}-k_1}}{sin\theta(1-cos\theta)^{I+\frac{1}{2}-k_1}}
\end{eqnarray}
and
\begin{eqnarray}
G^{k_1,k_2,I'_z}_{k_{1z},k_{2z}}(\alpha,\beta,\gamma,\alpha_I,\beta_I,\gamma_I)=
\sum_{mI_z}<k_1,m;I,I_z|k_2,k_{2z}>D^{k_1}_{m,k_{1z}}(\alpha,\beta,\gamma)D^I_{I_z,I'_z}(\alpha_I,\beta_I,\gamma_I)
\label{G}
\end{eqnarray}
Here $<k_1,m;I, I_z|k_2,k_{2z}>$ are the Clebsch-Gordon
coefficients, and
\begin{eqnarray}
D^{k_1}_{m,k}(\alpha,\beta,\gamma)=\langle k_1,m| e^{-i\alpha
S_z}e^{-i\beta S_y}e^{-i\gamma S_z} |k_1,k\rangle
\end{eqnarray}
is the standard $SU(2)$ representation matrix for the Euler angles
$\alpha,\beta,\gamma$ and $S_x,S_y,S_z$ are the standard $SU(2)$
spin matrixes in representation $k_1$.  Because the $SU(2)$ group
manifold and the three sphere $S^3$ are isomorphic, it can also be
viewed as the spherical harmonics on $S^3$.
$G^{k_1,k_2,I'_z}_{k_{1z},k_{2z}}$ forms a representation of the
$SO(4)$ group, which is the natural isometry group on the boundary
sphere $S^3$. It satisfies the following differential equation:
\begin{eqnarray}
& & \hat K_1^2 G^{k_1,k_2,I'_z}_{k_{1z},k_{2z}}=
k_1(k_1+1)G^{k_1,k_2,I'_z}_{k_{1z},k_{2z}},\ \
\hat K_2^2 G^{k_1,k_2,I'_z}_{k_{1z},k_{2z}} =k_2(k_2+1)G^{k_1,k_2,I'_z}_{k_{1z},k_{2z}}, \nonumber \\
& & \hat
I^2G^{k_1,k_2,I'_z}_{k_{1z},k_{2z}}=I(I+1)G^{k_1,k_2,I'_z}_{k_{1z},k_{2z}},
\ \
 \hat K_{1z} G^{k_1,k_2,I'_z}_{k_{1z},k_{2z}}= k_{1z} G^{k_1,k_2,I'_z}_{k_{1z},k_{2z}}, \ \
\hat  K_{2z}G^{k_1,k_2,I'_z}_{k_{1z},k_{2z}}= k_{2z}
G^{k_1,k_2,I'_z}_{k_{1z},k_{2z}},
 \label{so4}
\end{eqnarray}
where $\hat K_{1i}=\frac{1}{2}(\hat L_i+\hat P_i)$ and $\hat
K_{2i}= \frac{1}{2}(\hat L_i-\hat P_i)+\hat I_i$ are $SO(4)$
generators with $\hat L_{i}=-i\epsilon_{ijk}x_j\partial_k$ and
$\hat P_i=-i(x_4\partial_i-x_i\partial_4)$. The relationship
between the $<m>$ quantum numbers and the $SO(4)$ quantum numbers
$(k_1,k_2)$ are given by
\begin{eqnarray}
 k_1=\frac{m_3+m_4}{2},\ \ k_2=\frac{m_1+m_2}{2}.
 \label{km}
\end{eqnarray}
$G^{k_1,k_2,I'_z}_{k_{1z},k_{2z}}$, as given in the equation
(\ref{G}), is the most general solution to the differential
equations (\ref{so4}). The isospin direction can be normally
specified by two angles $\alpha_I$ and $\beta_I$. Here we followed
the trick first introduced in ref. \cite{instanton} which embeds
the $SU(2)$ isospin gauge group into a $SO(4)$ gauge group, as to
make the spatial and the isospin parts fully symmetric. This trick
is not necessary, but it
makes our discussion most general. At the end of our
calculations we will project back to the $\alpha_I$ and $\beta_I$
angles only. In this sense, $I'_z$ is simply a gauge index.
Different values of $I'_z$ solves the same differential equations
(\ref{so4}), and corresponds to the same physical state. If we
take $I'_z=I$ in equation (\ref{so4}), and using the
correspondence relation (\ref{km}), this single particle wave
function reduces exactly to the coherent states in ref.
\cite{4dqhe}. The function $f_{k_1,k_2}(\theta)$ localizes the
fermion on a given latitude, $x_5$, which is given by
$px_5=m_1+m_2-m_3-m_4$.

We now form the particle-hole operator
\begin{eqnarray}
\hat{\rho}(\theta,\alpha,\beta,\gamma,\alpha_I,\beta_I)&
= &
\sum_{<m>,<m'>}\bar\Psi^{\frac{m_1+m_2}{2},\frac{m_3+m_4}{2}}_{\frac{m_1-m_2}{2}\frac{m_3-m_4}{2}}
(\theta,\alpha,\beta,\gamma,\alpha_I,\beta_I,\gamma_I)
\nonumber \\
 & &
\Psi^{\frac{m'_1+m'_2}{2},\frac{m'_3+m'_4}{2}}_{\frac{m'_1-m'_2}{2}\frac{m'_3-m'_4}{2}}
(\theta,\alpha,\beta,\gamma,\alpha_I,\beta_I,\gamma_I)
c^+_{<m>}c_{<m'>} \label{p-h}
\end{eqnarray}
where $\Psi^{\frac{m_1+m_2}{2},\frac{m_3+m_4}{2}}_{\frac{m_1-m_2}{2}\frac{m_3-m_4}{2}}
(\theta,\alpha,\beta,\gamma,\alpha_I,\beta_I,\gamma_I)$ is the
single particle wave function given by Eq.(\ref{low}). Although the
single particle wave function depends on $\gamma_I$, $\hat{\rho}$
does not. This particle-hole operator is basically similar to (\ref{rho}),
the only difference is that the single particle eigenstates are
not ordinary plane waves. Just as in (\ref{rho}), we are going to perform a
series of transformations from the single particle coordinates to
center-of-mass and relative coordinates. Since $SO(4)=SU(2)\times
SU(2)$, the mathematical tools for accomplishing this task is the
standard angular momentum addition and decoupling. Because all
single particle eigenstates are also eigenstates of $x_5$, we
first transform into the center-of-mass and relative coordinates
$q$ and $n$ in the $x_5$ direction, by performing the following
substitution of variables,
\begin{eqnarray}
& &m_1=\frac{p}{4}+\frac{q}{2}+\frac{n}{2}+k_{1z}, \
m_2=\frac{p}{4}+\frac{q}{2}+\frac{n}{2}-k_{1z}, \
m_3=\frac{p}{4}-\frac{q}{2}-\frac{n}{2}+k_{2z}, \
m_4=\frac{p}{4}-\frac{q}{2}-\frac{n}{2}-k_{2z}, \ \nonumber \\
& & m'_1=\frac{p}{4}+\frac{q}{2}-\frac{n}{2}+k'_{1z}, \
m'_2=\frac{p}{4}+\frac{q}{2}-\frac{n}{2}-k'_{1z}, \
m'_3=\frac{p}{4}-\frac{q}{2}+\frac{n}{2}+k'_{2z}, \
m'_4=\frac{p}{4}-\frac{q}{2}+\frac{n}{2}-k'_{2z}. \nonumber
 \end{eqnarray}
where $k_{1z},k_{2z},k'_{1z}$ and $k'_{2z}$ are $SO(4)$ quantum
numbers.

The energy of the particle-hole pair is given by the
dipole separation along the $x_5$ direction.
$E_{<m>,<m'>}=2c\frac{n}{R}$. With this transformation,
the particle-hole operator now takes the following explicit form:
\begin{eqnarray}
&
&\hat{\rho}(\theta,\alpha,\beta,\gamma,\alpha_I,\beta_I)=
\sum_{n,q}g_{q,n}
(\theta)\sum_{k_{1z},k_{2z},k'_{1z},k'_{2z}} c^+_{p,q,n,k_{1z},k_{2z} }c_{p,q,n,k'_{1z},k'_{2z}}\nonumber  \\
 & &\sum_m <\frac{p}{4}-\frac{q+n}{2},m;\frac{p}{2},(k_{2z}-m)|\frac{p}{4}+\frac{q+n}{2},k_{2z}>
 D^{\frac{p}{4}-\frac{q+n}{2}}_{-m,-k_{1z}}(\alpha, \beta,\gamma)D^{\frac{p}{2}}_{-(k_{2z}-m),-s_z}(\alpha_I,\beta_I,\gamma_I)\ \nonumber \\
& &\sum_{m'}
<\frac{p}{4}-\frac{q-n}{2},m';\frac{p}{2},(k'_{2z}-m')|\frac{p}{4}+\frac{q-n}{2},k'_{2z}>
D^{\frac{p}{4}-\frac{q-n}{2}}_{m',k'_{1z}}(\alpha,
\beta,\gamma)D^{\frac{p}{2}}_{k'_2-m',s_z}(\alpha_I,\beta_I,\gamma_I)
\end{eqnarray}
where
 \begin{eqnarray}
 g_{q,n}(\theta)
&=&(-1)^{p}2^{-p}(p!)(p+2)^2(p+1)^2(\frac{p}{2}+1)sin^{-2}(\theta)(1+cos\theta)^{p+2}[\frac{1+cos\theta}{1-cos\theta}]^{q+1}
\nonumber \\
& & [\frac{1}
{(\frac{p}{2}-q-n)!(\frac{p}{2}-q+n)!(\frac{p}{2}+q+n+1)!
(\frac{p}{2}+q-n+1)!}]^{\frac{1}{2}} \nonumber
 \end{eqnarray}
and $s_z$ is a gauge index similar to $I_z'$ discussed earlier.
To obtain the particle-hole operator which is independent of $s_z$,
we define the projection:
 \begin{eqnarray}
 \hat{\rho}
 (\theta,\alpha,\beta,\gamma,\alpha_I,\beta_I)=\sum_{S}<\frac{p}{2},(-s_z);\frac{p}{2},s_z|S,0>\hat{\rho}^{S}(\theta,\alpha,\beta,\gamma,\alpha_I,\beta_I).
 \end{eqnarray}
Here $S$ is the iso-spin of the combined particle-hole pair and
$\hat{\rho}^S$ is independent of $s_z$. We now need to transform
the operators into a basis with well-defined center-of-mass
momentum along the boundary $S^3$. Let
$a_{p,q,n,T_1t_{1z},T_2t_{2z}}$ be particle-hole operator with
total $SO(4)$ quantum numbers $(T_1t_{1z}, T_2t_{2z})$,  which is
defined as
\begin{eqnarray} &
&a_{p,q,n,T_1t_{1z},T_2t_{2z}}=\sum_{k_{1z}k_{2z}k'_{1z}k'_{2z}}
<(\frac{p}{4}+\frac{q+n}{2}),-k_{2z};(\frac{p}{4}+\frac{q-n}{2}),k'_{2z}|T_2,t_{2z}> \nonumber \\
&
&<(\frac{p}{4}-\frac{q+n}{2}),-k_{1z};(\frac{p}{4}-\frac{q-n}{2}),k'_{1z}|T_1,t_{1z}>c^+_{p,q,n,k_{1z}k_{2z}}c_{p,q,n,k'_{1z}k'_{2z}}
\end{eqnarray}
The Clebsch-Gordon coefficients are only non-vanishing if $T_1\geq
n$ and $ T_2 \geq n$. Reversing the expansion, we obtain
\begin{eqnarray}
&
&c^+_{p,q,n,k_{1z}k_{2z}}c_{p,q,n,k'_{1z},k'_{2z}}=\sum_{T_1t_{1z},T_2t_{2z}}
 <T_2,t_{2z}|(\frac{p}{4}+\frac{q+n}{2}),-k_{2z};(\frac{p}{4}+\frac{q-n}{2}),k'_{2z}> \nonumber \\
&
&<T_1,t_{1z}|(\frac{p}{4}-\frac{q+n}{2}),-k_{1z};(\frac{p}{4}-\frac{q-n}{2}),k'_{1z}>a_{p,q,n,T_1t_{1z},T_2t_{2z}}.
\end{eqnarray}
Using this operator, we can simplify the particle-hole operator to
\begin{eqnarray}
\hat{\rho}^S(\theta,\alpha,\beta,\gamma,\alpha_I,\beta_I)=\sum_{T_1t_{1z},T_2t_{2z},n}
\hat{\rho}^S_{T_1t_{1z},T_2t_{2z},n}(\theta,\alpha,\beta,\gamma,\alpha_I,\beta_I)
\end{eqnarray}
where
\begin{eqnarray}
& &
\hat{\rho}^S_{T_1t_{1z},T_2t_{2z},n}(\theta,\alpha,\beta,\gamma,\alpha_I,\beta_I)=
(\frac{4\pi}{2S+1})^{\frac{1}{2}}G^{S}_{T_1t_{1z},T_2t_{2z}}
(\alpha,\beta,\gamma,\alpha_I,\beta_I)
\sum_q
g_{q,n}(\theta)\nonumber \\
& & \ \ \ \ \ \ \
((\frac{p}{4}-\frac{q+n}{2},\frac{p}{2})\frac{p}{4}+\frac{q+n}{2};
(\frac{p}{4}-\frac{q-n}{2},\frac{p}{2})\frac{p}{4}+\frac{q-n}{2};
(T_1,S)T_2)a_{p,q,n,T_1t_{1z},T_2t_{2z}}  \\
& &
 G^S_{T_1t_{1z},T_2t_{2z}}(\alpha,\beta,\gamma,\alpha_I,\beta_I) =
\sum_{m}<T_1,m;S,t_{2z}-m
|T_2,t_{2z}>D^{T_1}_{m,t_{1z}}(\alpha,\beta,\gamma)Y^S_{t_{2z},-m}(\beta_I,\alpha_I)
\label{Gs}
\end{eqnarray}
where $Y^S_{s_z}(\beta_I,\alpha_I)$ is the $SO(3)$ spherical
harmonics and $
((\frac{p}{4}-\frac{q+n}{2},\frac{p}{2})\frac{p}{4}+\frac{q+n}{2};
(\frac{p}{4}-\frac{q-n}{2},\frac{p}{2})\frac{p}{4}+\frac{q-n}{2};
(T_1,S)T_2)$ is the transformation coefficient  between two
coupling schemes of four angular momenta
$(\frac{p}{4}-\frac{q+n}{2},\frac{p}{2},\frac{p}{4}-\frac{q-n}{2},\frac{p}{2})$.
In the first coupling scheme, $(\frac{p}{4}-\frac{q+n}{2},
\frac{p}{2})$ couple to form $\frac{p}{4}+\frac{q+n}{2}$,
$(\frac{p}{4}-\frac{q-n}{2}, \frac{p}{2})$ couple to form
$\frac{p}{4}+\frac{q-n}{2}$ and $(\frac{p}{4}+\frac{q+n}{2},
\frac{p}{4}+\frac{q-n}{2})$ couple to form $T_{2}$. In the second
coupling scheme, $(\frac{p}{4}-\frac{q+n}{2},
\frac{p}{4}-\frac{q-n}{2})$ couple to form $T_1$, $(\frac{p}{2},
\frac{p}{2})$ couple to form $S$ and $(T_1, S)$ couple to form
$T_{2}$.
 The
following formula is used in above derivation,
\begin{eqnarray}
<j,n;j',n'|L,(n+n')>D^L_{M,n+n'}(A)=\sum_{m,m'}<j,m;j',m'|L,M>D^j_{mn}(A)D^{j'}_{m'n'}(A).
\label{f1}
\end{eqnarray}
The transformation coefficient can be explicitly written in terms
of the  $9-j$ symbol\cite{angular9j},
\begin{eqnarray}
& &
((\frac{p}{4}-\frac{q+n}{2},\frac{p}{2})\frac{p}{4}+\frac{q+n}{2};
(\frac{p}{4}-\frac{q-n}{2},\frac{p}{2})\frac{p}{4}+\frac{q-n}{2};
(T_1,S)T_2) \nonumber \\
& &
=[(\frac{p}{2}+q+n)(\frac{p}{2}+q-n)(2T_1+1)(2S+1)]^{\frac{1}{2}}\left\{
\begin{array}{ccc}
  \frac{p}{4}-\frac{q+n}{2}& \frac{p}{2} & \frac{p}{4}+\frac{q+n}{2} \\
   \frac{p}{4}-\frac{q-n}{2}& \frac{p}{2} & \frac{p}{4}+\frac{q-n}{2} \\
  T_1 & S & T_2 \\
\end{array} \right\}.
\end{eqnarray}
$G^{S}_{T_1t_{1z},T_2t_{2z}}$ is the exact eigen-function
with $SO(4)$ quantum numbers $(T_1,T_2)$, where the $SO(4)$
generators are defined in terms of the center of mass coordinates of the
particle and the hole,  $\hat T_1=\frac{\hat L+ \hat P}{2}$ and $\hat
T_2=\frac{\hat L-\hat P}{2}+\hat S$. Therefore, it satisfies the
following equations
\begin{eqnarray}
\label{eqso4}
 & &(\hat{L}^2+\hat{P}^2)) G^{S}_{T_1t_{1z},T_2t_{2z}}=
4T_1(T_1+1)G^{S}_{T_1t_{1z},T_2t_{2z}}
\\
& & (\hat{L}\cdot \hat S -\hat P \cdot \hat S )
G^{S}_{T_1t_{1z},T_2t_{2z}}=
[T_2(T_2+1)-T_1(T_1+1)-S(S+1)]G^{S}_{T_1t_{1z},T_2t_{2z}}.
\label{eqso40}
\end{eqnarray}
At the boundary, we take a fixed value for $\theta$. In the limit of large $p$, using the
asympotic formula of the 9-j symbol, one can show that  up to a
constant factor,
\begin{eqnarray}
\hat{\rho}^S_{T_1t_{1z},T_2t_{2z},n}(\theta,\alpha,\beta,\gamma,\alpha_I,\beta_I)
=
G^S_{T_1t_{1z},T_2t_{2z}}(\alpha,\beta,\gamma,\alpha_I,\beta_I)e^{-l_0^2(\frac{n}{R})^2}\hat
b_{n,T_1t_{1z},T_2t_{2z}} \label{operator}
\end{eqnarray}
where $\hat b_{n,T_1t_{1z},T_2t_{2z}}=\sum_q
e^{-l_0^2(\frac{q}{R})^2}\hat a_{p,q,n,T_1t_{1z},T_2t_{2z}}$.
%

We are now in a position to define the
concept of {\it extremal dipole operators} within the operators
defined in (\ref{operator}). In general, $T_1\geq n$ and $T_2 \geq
n$. Extremal dipole are these ones for which
\begin{eqnarray}
T_2=T_1-S=n \ \  or \ \ \ T_2-S=T_1=n
\label{extreme}
\end{eqnarray}
with $S=0, 1, 2,... $. For a given pair $(T_1,T_2)$, we choose the
smallest possible value of $S=|T_1-T_2|$. $(T_1,T_2)$ are the
physically observable quantum numbers, different possible values of $S$
for a given pair $(T_1,T_2)$ represent the same helicity state.
Since $(T_1,T_2)$ is
basically the momentum ${\bf q}$ along the three dimensional
boundary, and $n$ is the dipole distance along the extra fourth
dimension, these operators have a well defined relationship
between the momentum ${\bf q}$ and the dipole distance $\Delta
x_5=|{\bf q}|l_0^2$. This is the maximally allowed value of the
dipole moment for fixed ${\bf q}$. The time evolution of these
operators are given by the quantum mechanical Heisenberg equation
of motion
\begin{eqnarray}
\frac{\partial}{\partial t} \hat{\rho}^S_{T_1t_{1z},T_2t_{2z},n} =
i [H, \hat{\rho}^S_{T_1t_{1z},T_2t_{2z},n}] =-i\frac{2cn}{R}
\hat{\rho}^S_{T_1t_{1z},T_2t_{2z},n} \label{time}
\end{eqnarray}
Equations (\ref{eqso4}, \ref{eqso40}), (\ref{extreme}) and (\ref{time}) are the
desired relativistic equations on $S^3$.

One can also show explicitly that the extremal dipole constructed
above are well-localized in the three dimensional coordinates
$(\alpha,\beta,\gamma)$ of the particle and the hole.
The localization length is determined by the magnetic length
$l_0$. To prove this statement, let the coordinates of the particle be
$(\alpha,\beta,\gamma)$ and the coordinates of the hole be
$(\alpha+\Delta \alpha, \beta +\Delta \beta,\gamma +\Delta
\gamma)$.  The wave function of state $(T_1t_{1z}, T_2t_{2z})$ now
also depends on the relative angles $(\Delta \alpha, \Delta
\beta,\Delta \gamma)$ between particle and hole, which is
explicitly given by  $ D^{\frac{p}{4}-\frac{q+n}{2}}_{mm'}(\Delta
\alpha, \Delta \beta,\Delta \gamma)$. In the limit of
$R\rightarrow \infty$, the amplitude is determined by the diagonal
term of the matrix which is proportional to
$e^{-\frac{p}{8}(\Delta\beta)^2} \propto e^{-\frac{\Delta
X^2}{4l_0^2}}$, where $\Delta X$ is relative distance between
particle  and hole in flat space. For the extremal dipole states,
the separation between the particle
and the hole in the $X_5$ coordinate is given by $|{\bf q}|l_0^2$, where the
linear momentum of the pair is ${\bf q}$ in the three dimensional
flat space.

We now show that they reduce to the familiar massless relativistic
equations in the flat space limit, where $R\rightarrow \infty$, and
the three sphere becomes the three dimensional Euclidean space. We
take flat space limit on the equations (\ref{eqso4},
\ref{eqso40}). For the extreme dipole operators, we define the
wave functions $\psi_{S,+}(x,t)$ and $\psi_{S,-}(x,t)$ to be  the
flat space limit of $G^{S}_{n+Sm_z;nn_z}$ and
$G^{S}_{nm_z;n+Sn_z}$ respectively. In the flat space limit, we
expand the operators around the north pole point $(0,0,0,R)$.  Compared with
the eigenvalues of the
momentum operator $\hat P$ and $T_1, T_2$, which are the
order of $R$, the angular momentum operators $\hat L_i$  and the
fixed iso-spin value  $S$  are of the order of one. Therefore, $\hat L_i$
and $S(S+1)$ vanish in this limit. Equations (\ref{eqso4},
\ref{eqso40}) then become
\begin{eqnarray}
& &\hat P^2\psi_{S,\pm}(x,t)=\frac{\hat
E^2}{c^2}\psi_{S,\pm}(x,t),
\\
& &  \frac{\hat S}{S}\cdot \hat P\psi_{S,\pm}(x,t) = \pm \hat E
\psi_{S,\pm}(x,t)=\pm i\partial_t \psi_{S,\pm}(x,t).
\label{chiral}
\end{eqnarray}
These are exactly the massless relativistic wave equations with
helicity $S$. The operator $\hat S$ was originally introduced as
a differential operator with respect to $\alpha_I$ and $\beta_I$.
But for a given $S$, it can also be implemented as a
$(2S+1)\times (2S+1)$ matrix, acting on a $(2S+1)$ component
tensor field.

Now we show that these two equations together are equivalent to
the Maxwell equation in the case of  $S=1$ and the linearized
Einstein equation in the case of $S=2$. When $S=1$,
$\psi_{S,\pm}(x,t)$ is a vector denoted by
$\phi^{\pm}_{\mu}(x,t)$, which satisfies $\hat S_{\sigma}
\phi^{\pm}_{\mu}(x,t) =
i\epsilon_{\mu\sigma\alpha}\phi^{\pm}_{\alpha}(x,t)$. In the case
of $S=2$, $\psi_{S,\pm}(x,t)$ is a rank-two symmetric traceless
tensor denoted by $\phi^{\pm}_{\mu\nu}(x,t)$, which satisfies
$\hat S_{\sigma} \phi^{\pm}_{\mu\nu}(x,t) =
i\epsilon_{\mu\sigma\alpha}\phi^{\pm}_{\alpha\nu}(x,t)+
i\epsilon_{\nu\sigma\alpha}\phi^{\pm}_{\mu\alpha}(x,t)$. Thus they
satisfy
 \begin{eqnarray}
 [(\hat S\cdot \hat P)^2-\hat P^2]\phi_{\mu}=0, \ \  [(\hat S\cdot \hat P)^2-4 \hat P^2]\phi_{\mu\nu}=0.
 \end{eqnarray}
Above equations can be simplified to the following form,
 \begin{eqnarray}
 \partial_{\mu}\nabla\cdot\phi^{\pm}=0, \ \  \partial_{\mu}(\partial_{\gamma}
 \phi^{\pm}_{\gamma\nu})
 +\partial_{\nu}(\partial_{\gamma}\phi^{\pm}_{\gamma\mu})
-\frac{2}{3}\delta_{\mu\nu}\partial_{\gamma}\partial_{\sigma}\phi^{\pm}_{\gamma\sigma}
 =0.
 \end{eqnarray}
Since there is no constant source, the above two equations are equivalent to
\begin{eqnarray}
\nabla\cdot\phi^{\pm}=0, \  \
\partial_{\alpha}\phi^{\pm}_{\alpha\mu}=0. \label{divergence}
\end{eqnarray}
Together with Eq.\ref{chiral}, which can be explicitly written for by
$\phi^{\pm}_{\mu}$ and $\phi^{\pm}_{\mu\nu}$ as
\begin{eqnarray}
\epsilon_{\mu\alpha\beta}\partial_{\alpha}\phi^{\pm}_{\beta}=
\pm\frac{i}{c}\partial_t\phi^{\pm}_{\mu}, \ \
\epsilon_{\mu\alpha\beta}\partial_{\alpha}\phi^{\pm}_{\beta\nu}+
\epsilon_{\nu\alpha\beta}\partial_{\alpha}\phi^{\pm}_{\mu\beta}=\pm
2\frac{i}{c}\partial_t\phi^{\pm}_{\mu\nu}, \label{e2} \label{curl}
\end{eqnarray}
the above two equations give the complete free Maxwell equation and
linearized Einstein equation. $\phi^\pm_\mu$ is nothing but the linear
combination $E_\mu \pm i H_\mu$ of the electric and the magnetic field.
We can also introduce a vector
potential $A_{\mu}$  and a symmetric tensor potential $h_{\mu\nu}$
respectively to describe Maxwell and linearized Einstein equation.
The explicit relations between them are given by
\begin{eqnarray}
& &\phi^{\pm}_{\mu}=-\frac{1}{c}\partial_t A_{\mu}\pm
i\epsilon_{\mu\alpha\beta}\partial_{\alpha}A_{\beta} \\
& & \phi^{\pm}_{\mu\nu}=-\frac{1}{c}\partial_t h_{\mu\nu}\pm
\frac{i}{2}(\epsilon_{\mu\alpha\beta}\partial_{\alpha}h_{\beta\nu}+
\epsilon_{\nu\alpha\beta}\partial_{\alpha}h_{\mu\beta}).
\label{Maxwell-Einstein}
\end{eqnarray}

We have now shown that the extremal dipole operators
are local in space and satisfy massless relativistic equations
with well defined helicities. In this precise sense, both the
Maxwell and the Einstein equations (\ref{Maxwell-Einstein}) have
been derived as operator equations of motion from a
single non-relativistic Hamiltonian (\ref{H}), with single
particle eigenstates given by (\ref{low}).

After proving this exact mathematical result we now make some
physical observations, and give a critical discussion of what lies
ahead.

1) The extremal dipole operators can be naturally identified
with operators which create shape distortions at the boundary of
the incompressible 4D QHE droplet. The equilibrium shape of the droplet is a
perfect sphere $S^3$. However, this sphere is composed of $2I+1$
different copies, one for each isospin direction $\hat n$, all
with exactly the same radius $x^F_5(\alpha,\beta,\gamma;\hat
n)=x^F_5$. This is somewhat similar to the fermi surface of
electrons, which has two different sheets $k_F(\hat
k,\sigma=\pm)$ for up and down spins. Once the droplet shape is
distorted, every isospin can have its own, and in general
different, shape of the surface. Therefore, there are in general
$2I+1$ different collective isospin modes for a given spatial
harmonics of the distortion. The scalar mode is created by the
extremal dipole operator with $S=0$, which uniformly averages over
all different isospin sheets $x^F_5(\alpha,\beta,\gamma;\hat n)$
at a given spatial position $(\alpha,\beta,\gamma)$. The $S=1,2$
modes single out the dipolar and quadrupolar distortions of the
different isospin sheets. If amplitudes for all the different
helicity modes are obtained, the shape of the droplet
$x^F_5(\alpha,\beta,\gamma;\hat n)$ for every isospin direction
$\hat n$ can be reconstructed exactly. In this sense, the extremal
dipole operators create the hydrodynamical modes of the droplet.

Since extreme dipole operators are local in space, one can define their
correlation functions and the imaginary parts contain only $\delta$ function
peaks. However, considered as part of the full density correlation
function, their energy lies on the upper edge of the
continuum. The continuum also has contributions from the non-extremal
dipoles. These non-extremal dipoles are best described in terms of the original
fermions. If one turns on a repulsive interaction among the
fermions, it will lead to an attractive force between the particle and
the hole of the dipole.
Since the extremal dipole pairs are maximally localized already in their
relative coordinates, one expects that they will be further {\it stabilized}
by interactions, similar to the zero sound mode of the fermi liquid.
In general we expect a rich phase diagram of possible phases in this
model. Among the possible phases are liquid states where
a full or partial energy gap opens up in the fermionic part of the spectrum.
According to our experience with superconductivity, the collective modes
is expected to be unaffected. In this case, the
collective modes found in this work are well separated from the fermionic
continuum, and we can construct an effective theory for these
bosonic collective modes.

2) Bosonic particles occur with both helicities. This is different
from the edge states of the 2D QHE droplet, which are
chiral\cite{halperin,wen,stone,fisher}. This fact can be understood through
the discrete symmetries of the model. The $SU(2)$ monopole field imposes
an isospin-orbit coupling, of the type ${\bf L}\cdot{\bf S}$, which preserves
time reversal symmetry $T$. Three dimensional parity operation $P$ can
be defined as an interchange of the $SO(4)$ quantum numbers $(k_1,k_2)$,
therefore, the fermionic states generally break $P$. One can also define
a charge conjugation operation $C$, which interchanges a particle with a
hole. If the droplet is filled up to the equator, $CP$ is an exact
symmetry of the Hamiltonian. In general, this is an excellent symmetry
when only states close to the droplet surface are considered. The bosonic dipole
states are formed from particle hole pairs, the charge conjugation operation
acts trivially on the pair. Therefore, these states have to form representation
of parity $P$, which explains why both helicity states occur. Parity violating
effects can only be observed for operators with non-zero fermion number.

3) The most non-trivial feature of the theory is the helicity.
It is known
from the representation theory of the Poincare group that massless relativistic
particles only form representations of the $U(1)$ helicity group, but not
the spin $SU(2)$ group. This feature is very hard to produce in an
ordinary non-relativistic system. In this model, particles carry $SU(2)$
isospin labels, but the independent isospin rotation is not a symmetry of the
Hamiltonian. Only a combined isospin and space rotation is a symmetry of the
Hamiltonian. It is this property which enables the extremal dipole states
to have exactly the same symmetry as the massless relativistic particles
with non-trivial helicities.

4) There is another way in which we can view the different branches of the
hydrodynamical modes. As mentioned in ref.\cite{4dqhe}, the total configuration
space of 4D QHE is locally $S^4\times S^2$. The configuration space at the droplet
boundary is $S^3\times S^2$. Viewed from this five dimensional configuration
space, there is only a single scalar hydrodynamical mode. However, when projected onto
the three dimensional base space, different modes on the iso-spin space
$S^2$ appear as different branches in the base three dimensional space.
Originally, $S^2$ was introduced as a isospin degree of freedom over $S^4$,
however, the unit tangent bundle of the boundary surface $S^3$ is also
$S^2$. Therefore, different modes of distortion on the isospin sphere $S^2$
can be naturally identified with the spin degree of freedom at the boundary.

5) Since the dimension of the total configuration space is higher than the
dimension of the base space, this theory bears similarities to the Kaluza-Klein
theory, but with two important differences. First, the total configuration
space is a topologically non-trivial fiber bundle. Second, the iso-spin space
does not have a small radius. This leads to the ``embarrassment of riches" problem
mentioned in \cite{4dqhe}. In order to solve this problem, we need to
find a mechanism where higher isospin states obtain mass gaps dynamically,
through interactions. This way, the low energy degrees of freedom would scale
correctly with the dimension of the base space.

In condensed matter physics, there are actual examples where this type of
phenomena occurs. Consider a valance bond solid state, where higher spin
degrees of freedom $S$ reside on lattice points with coordination number
$Z$\cite{aklt}. A higher spin
degree of freedom can be viewed as a symmetrized product of $2S$ spin $1/2$
objects. In a valance bond configuration, most of the spin $1/2$ degrees
of freedom are ``contracted" with the other spin degrees of freedom on the
neighboring sites to form spin singlets. In the valence bond solid ground
state, there are only $2S-nZ$ effective spin $1/2$ degrees of freedom left
on each site, where $n$ is the largest integer such that $2S-nZ$ is non-negative.
Therefore, while a non-interacting system can have an arbitrarily large spin
degree of freedom $2S$ on each site, the strong coupling
fix point only has a small effective spin degree of freedom on each site.
The small spin degree of freedom is separated from the higher spin excitations
by finite energy gaps. By a similar mechanism of forming spin singlets,
the effective spin degree of freedom can be lowered in our model.

This comment applies in particular to the fermionic
states, which are not well understood in the current version of the theory.
Since they carry large iso-spin quantum
numbers, they can not be identified with any familiar relativistic particles.
In order to obtain a sensible low energy theory with
a full relativistic spectrum, one needs to find a mechanism so that the
fermionic states become fully or partially gapped, while leaving the
collective modes unaffected. This is exactly what happens when a fermionic
system becomes superconducting. Spin gap mechanisms mentioned above could
also be a possibility here. It is important to identify all strong
coupling fix points of the system, and identify the fermionic
spectrum at these fix points. Interesting strong coupling fix points are
those where higher spin states have higher energy gaps.

5) The underlying mathematical structure of the current approach is the
non-commutative geometry\cite{connes} defined by Eq. (\ref{non-commutation}).
Unlike previous approaches\cite{szabo}, this relation treats all four
Euclidean dimensions on equal footing. If we interpret $X_4$ as energy,
which is dual to time, this quantization rule seem to connect space,
time, spin and the fundamental length unit $l_0$ in an unified fashion.
In the lowest $SO(5)$ level, there is no ordinary non-relativistic
kinetic energy. All the single particle states are representations of this algebra.
The non-trivial features identified in this work all have their roots in this
algebra.

6) This work may have many connections with related ideas.
Our approach is motivated by
the idea of ``emergence"\cite{bob}, and could in particular be related
to Volovik's approach
based on momentum space topology\cite{volovik1,volovik2}.
The problem of higher spin massless particles has been investigated extensively
in field theory.
Recently, an algebraic structure of non-commutative geometry has been
identified for this problem\cite{vasiliev}. It would be interesting
to investigate its connection to our work. The general connection
between quantum Hall effect and the brane solutions of the string theory\cite{lenny}
is also worth exploring for our 4D QHE model.

We would like to thank Prof. B.J. Bjorken, S. Dimopoulos, M. Freedman,
D. Gross, R.B. Laughlin, A. Ludwig, C. Nayak, J. Polchinsky, A. Polyakov,
L. Susskind and G. Volovik for many stimulating discussions.
This work is supported by the NSF under grant numbers DMR-9814289.

\newpage

{\bf Appendix}

We have shown that the Heisenberg equation of motion for the extremal
dipole operators satisfy relativistic wave equations with non-trivial
helicities. It is also possible to show directly that these operators
reduce exactly to the solutions of relativistic wave equations in the
flat space limit. The mathematical tools needed for this demonstration
is called the contraction limit of the $SO(4)$ group when the representations
are large, and these tools are provided in ref. \cite{talman,itzykson}.

We shall show that the extremal dipole wave functions given in (\ref{Gs}),
$G^{S}_{n,t_{1z};n+S,t_{2z}}(\alpha,\beta,\gamma,\alpha_I,\beta_I)$ and
$G^{S}_{n+S,t_{1z};n,t_{2z}}(\alpha,\beta,\gamma,\alpha_I,\beta_I)$,
are the wave functions of particles with helicities
$\pm S$ in the flat space limit. In the following, a normalization
factor is added to the definition of the wave functions, i.e.,  we
define,
\begin{eqnarray}
G^S_{T_1t_{1z},T_2t_{2z}}(\alpha,\beta,\gamma,\alpha_I,\beta_I)
=\frac{(2T_1+1)^{\frac{1}{2}}}{\sqrt{2}\pi}
\sum_{m}<T_1,m;S,t_{2z}-m
|T_2,t_{2z}>D^{T_1}_{m,t_{1z}}(\alpha,\beta,\gamma)Y^S_{t_{2z}-m}(\beta_I,\alpha_I).
\end{eqnarray}

First, we consider  $S=0$. In this case, $G^0_{n
t_{1z},nt_{2z}}(\alpha,\beta,\gamma)$ only depends on the spatial
coordinates. The coordinate space for $\alpha,\beta$ and $\gamma$
is $S^3$ which is isomorphic to the $SU(2)$ group manifold.
Let $V$ be an element in $SU(2)$ group. $V$ can be parameterized
as
\begin{eqnarray}
  V=x_4I+ix_i\sigma_i=\left( \begin{array}{ll}
             x_4+ix_3 & -x_2+ix_1 \\
             x_2+ix_1 & x_4-ix_3
             \end{array} \right),
\end{eqnarray}
which defines a one to one mapping from $SU(2)$ group to $S^3$.
We define $(R,R' )$ to be a pair of elements in $SU(2)$,  which
creates
the following rotation on $SU(2)$ group, $V'=RVR^{'-1}$. The whole
set of pairs $(R,R')$ forms a $SO(4)$ group defined in terms of above
operations. The subgroup $(R,R)$ leaves $x_4$ invariant rotation.
It describes the $SO(3)$ rotation group in space $x_1, x_2$ and
$x_3$. Let $g(\psi)= cos(\psi)+isin(\psi)\sigma_z$ denote a
special element of $SU(2)$,  which defines a
rotation by an angle $\psi$ in both of $x_3-x_4$ and $x_1-x_2$
plane.  Then,  any $SU(2)$ elements, $V$, can be generated by
performing an rotation $(R,R)$ on $g(\psi)$, {\it i.e.} $V$ can be
decomposed into the form  $V=R g(\psi)\Gamma R^{-1}$, where $ \Gamma= \left(
\begin{array}{ll}
              0 &1\\
              -1&0
              \end{array} \right) $,
              is chosen for convenience.
              Thus,
\begin{eqnarray}
 G^0_{nt_{1z},nt_{2z}}(V)= \frac{(2l+1)^{\frac{1}{2}}}{\sqrt{2}\pi}
 \sum_{s_1,s_2} D^{n}_{t_{2z}s_1}(R)
 D^n_{s_1s_2}(g(\psi)\Gamma)D^n_{s_2t_{1z}}(R^{-1}).
 \end{eqnarray}
 Since the total angular momentum in three
 dimensional flat space is a good quantum number, we define a set of basis
wave functions $G^{0}_{JM,n}(V)$,
\begin{eqnarray}
G^{0}_{JM,n}(V) = \sum_{t_{1z},t_{2z}}
<n,t_{1z};n,t_{2z}|JM>G^0_{nt_{1z},nt_{2z}}(V) ,
\end{eqnarray}
Applying  equation (\ref{f1}), we obtain
\begin{eqnarray}
G^{0}_{JM,n}(V) =\frac{(2n+1)^{\frac{1}{2}}}{\sqrt{2}\pi}
\sum_{M't_{1z}t_{2z}} <n,s_1;n,s_2
|JM'>D^{J}_{MM'}(R)D^n_{s_1s_2}(g(\psi)\Gamma).
\end{eqnarray}
The $D(g(\psi)\Gamma)$ matrix has a very simple form,
\begin{eqnarray}
 D^n_{s_1s_2}(g(\psi)\Gamma)=(-1)^{n-s_1}\delta_{s_1,-s_2}exp(-2is_1\psi).
 \end{eqnarray}
Thus, we obtain the following result,
\begin{eqnarray}
G^{0}_{JM,n}(V) =
\frac{(2n+1)^{\frac{1}{2}}(2J+1)^{\frac{1}{2}}}{\sqrt{2}\pi}i^J
H_{n,J}(\psi)Y^J_M(\theta,\phi)
\end{eqnarray}
where
\begin{eqnarray}
H_{n,J}(\psi) =\frac{1}{(2J+1)^{\frac{1}{2}}i^J} \sum_{s}
(-1)^{n-s}<n,s;n,-s|J0>exp(-2is\psi).
\end{eqnarray}
and  $\theta, \phi$ are corresponding coordinates  of the rotation
$(R,R)$ parametrized in $(x_1,x_2,x_3)$ space.
From equation (\ref{bessel}) below, we can see that $G^0$ reduces to the
usual solution of the scalar equation in the spherical coordinate
system of the flat space.

Now we consider arbitrary helicity values $S$. We define the
following wave functions,
\begin{eqnarray}
G^{S,+}_{JM,n}(V,\beta_I,\alpha_I) = \sum_{t_{1z}t_{2z}} <n+S,
t_{1z};n,t_{2z}|JM> G^{S}_{n+S,
t_{1z};n,t_{2z}}(V,\beta_I,\alpha_I).
\end{eqnarray}
Using $6-j$ symbol which involves with the sum of three angular
momentum $(n, n, S)$, we can readily obtain
\begin{eqnarray}
G^{S,+}_{JM,n}(V,\beta_I,\alpha_I) &=& \sum_{Ls_1s_2}
(-1)^{2n+S+J}(2L+1)^{\frac{1}{2}}(2n+2S+1)^{\frac{1}{2}} \left\{
\begin{array}{lll}
 n&n&L \\
 S&J&n+S
 \end{array} \right\}  \nonumber \\
 & & <L,s_1;S,s_2|JM>G^{0}_{Ls_1,n}(V) Y^S_{s_2}(\beta_I,\alpha_I) \nonumber \\
&=& \sum_{L}(\frac{2}{\pi})^{\frac{1}{2}}i^L
(-1)^{2n+S+J}(2L+1)^{\frac{1}{2}}(2n+2S+1)^{\frac{1}{2}} \left\{
\begin{array}{lll}
 n&n&L \\
 S&J&n+S
 \end{array} \right\}  \nonumber \\
 & &H_{n,L}(\psi)
 (\sum_{s_1s_2}<Ls_1;Ss_2|JM>Y^L_{s_1}(\theta,\phi)Y^S_{s_2}(\beta_I,\alpha_I)
\end{eqnarray}

Taking the flat space limit,($n \rightarrow \infty$ and $
\psi\rightarrow \frac{-px}{2n}$),
\begin{eqnarray}
\left\{
\begin{array}{lll}
 n&n&L \\
 S&J&n+S
 \end{array} \right\} = \left\{
\begin{array}{lll}
 J&S&L \\
 n&n&n+S
 \end{array} \right\} \rightarrow \frac{(-1)^{S+n}}{(2n)^{1/2}} \left(
\begin{array}{lll}
 J&S&L \\
 S&-S&0
 \end{array} \right)
\end{eqnarray}
\begin{eqnarray}
(2n)^{-1/2}H_{n,J}(\frac{-px}{2n}) \rightarrow j_J(px)
\label{bessel}
\end{eqnarray}
where $j_J(x)$ is the spherical Bessel function of order $J$,
defined as
\begin{eqnarray}
j_J(x) = (-x)^J(\frac{1}{x}\frac{d}{dx})^J(\frac{sinx}{x})
\end{eqnarray}
Therefore, in the flat limit, up to a constant normalization factor, we
obtain
\begin{eqnarray}
\label{flat function}
 G^{S,+}_{JM,n}(V,\beta_I,\alpha_I) &\rightarrow&
\frac{1}{(2J+1)^{\frac{1}{2}}}\sum_{L} i^L(2L+1)^{\frac{1}{2}}
 <S,S;L0|J,S> j_L(px) \Psi^{JM}_{LS}(\theta,\phi,\beta_I,\alpha_I)
 \end{eqnarray}
 where
 \begin{eqnarray}
 \Psi^{JM}_{LS}(\theta,\phi,\theta_I,\phi_I)=\sum_{ls}<L,l;S,s|JM>Y^L_{l}(\theta,\phi)Y^S_s(\beta_I,\alpha_I)
\end{eqnarray}
The right side of Eq. (\ref{flat function}) is exactly the wave
function of a bosonic particle with helicity
$S$, momentum $p$, and total angular momentum $(J,M)$. The
derivation for helicity $-S$ wave function $G^{S,-}_{JM,n}(V,\beta_I,\alpha_I)$
follows the same procedure. For $S=\pm 1$, the wave function reduces to the
usual expansion of the Maxwell field in the spherical
coordinate system.

\bibliographystyle{prsty}
\bibliography{dipole}

\begin{thebibliography}{10}

\bibitem{laughlin}
R.~B. Laughlin, Phys. Rev. Lett. {\bf 50},  1395  (1983).

\bibitem{4dqhe}
S.-C. Zhang and J.-P. Hu, Science {\bf 294},  823  (2001).

\bibitem{yang1}
C.~N. Yang, J. Math. Phys. {\bf 19},  320  (1978).

\bibitem{yang2}
C.~N. Yang, J. Math. Phys. {\bf 19},  2622  (1978).

\bibitem{jackiw}
R. Jackiw and C. Rebbi, Phys. Rev. D {\bf 14},  517  (1976).

\bibitem{polyakov}
A. Belavin and A. Polyakov, Nucl. Phys. B {\bf 123},  429  (1977).

\bibitem{instanton}
A. Belavin, A. Polyakov, A. Schwartz, and Y. Tyupkin, Phys. Lett. B {\bf 59},
  85  (1975).

\bibitem{halperin}
B.~I. Halperin, Phys. Rev. B {\bf 25},  2185  (1982).

\bibitem{wen}
X.~G. Wen, Phys. Rev. Lett. {\bf 64},  2206  (1990).

\bibitem{stone}
M. Stone, Phys. Rev. B {\bf 42},  8399  (1990).

\bibitem{fisher}
K. Moon {\it et~al.}, Phys. Rev. Lett. {\bf 71},  4381  (1993).

\bibitem{weinberg-witten}
S. Weinberg and E. Witten, Phys. Lett. B {\bf 96},  59  (1980).

\bibitem{kallin}
C. Kallin and B. Halperin, Phys. Rev. B {\bf 30},  5655  (1984).

\bibitem{haldane}
V. Pasquier and F. Haldane, Nucl. Phys. B {\bf 516},    (1998).

\bibitem{dhlee}
D. Lee, Phys. Rev. B {\bf 60},  5636  (1999).

\bibitem{angular9j}
A.~R. Edmonds, {\em Angular Momentum in Quantum Mechanics} (Academic Press, 1957).

\bibitem{aklt}
I. Affleck, T. Kennedy, E.~H. Lieb, and H. Tasaki, Phys. Rev. Lett. {\bf 59},
  799  (1987).

\bibitem{connes}
A. Connes, {\em Noncommutative Geometry} (Academic Press, 1994).

\bibitem{szabo}
R. Szabo, hep-th/0109162  .

\bibitem{bob}
R. Laughlin and D. Pines, Proc. Natl. Acad. Sc. USA {\bf 97},  28  (2000).

\bibitem{volovik1}
G. Volovik, Phys. Rep. {\bf 351},  195  (2001).

\bibitem{volovik2}
G. Volovik, gr-qc/0112016  .

\bibitem{vasiliev}
M.~A. Vasiliev, hep-th/9910096  .

\bibitem{lenny}
L. Susskind, hep-th/0101029  .

\bibitem{talman}
J. Talman, {\em Special Functions, A Group Theoretic Approach} (Benjamin Inc,
  Boston, MA, 1968).

\bibitem{itzykson}
M. Bander and C. Itzykson, Rev. Modern Phys. {\bf 38},  330  (1966).

\end{thebibliography}

\end{document}